# KL-Mat : Fair Recommender System via Information Geometry

Hao Wang

haow85@live.com

Ratidar.com

*Abstract*— Recommender system has intrinsic problems such as sparsity and fairness. Although it has been widely adopted for the past decades, research on fairness of recommendation algorithms has been largely neglected until recently. One important paradigm for resolving the issue is regularization. However, researchers have not been able to come up with a consensusly agreed regularization term like regularization framework in other fields such as Lasso or Ridge Regression. In this paper, we borrow concepts from information geometry and propose a new regularization-based fair algorithm called KL-Mat. The algorithmic technique leads to a more robust performance in accuracy performance such as MAE. More importantly, the algorithm produces much fairer results than vanilla matrix factorization approach. KL-Mat is fast, easy-to-implement and explainable.

## I. Introduction

Recommender system is one of the major research focus in the machine learning community. Since its initial application in e-commerce websites such as Amazon, it has drawn attention from tens of thousands of researchers and industrial practitioners worldwide. Annual research venues such as KDD, SIGIR, ICML and RecSys receive a large number of manuscripts on the topic. The utmost important task for recommender system research is to increase its performance in the real world environment, measured in MAE, Precision@K or nDCG, etc.

Since most people focus on lifting the accuracy performance of recommender system, other aspects of recommender systems have been largely ignored until recent years. As we all know in our social environment, power law is a dominating effect in social networks, city population distribution, human word occurence distributions, etc. Recommender system is also susceptible to the power law phenomenon and the problem is known as the fairness problem.

Google researcher Ed Chi [1] introduces one of the first algorithmic designs to tackle the fairness problem. He names the algorithm Focused Learning. Since the introduction, many researchers have delved into the area and proposed many regularization terms [2][3][4] in the years that follows. However, these technqiues look more like engineering tricks rather than scientific invention, and there has been no direct attack on the mathematically accurate regularization technique until 2021.

In the year of 2021, Hao Wang [5] comes up with a regularized matrix factorization approach named Zipf Matrix Factorization. The researcher uses a statistical estimator that approximates the exponent of the item popularity distribution as power law distribution. For the first time in the research history of recommender system, someone has introduced rigorous math into selection of regularization term.

The impact of appropriate selection of regularization can never be overemphasized. For example, in the area of health recommendation, if the most popular doctors get recommended exponentially more frequently than others, it's going to cause problems. Even worse, if simple symptoms such as cold or fever dominates more severe but rarer diseases, it's going to diminish the opportunities for patients with more grave diseases.

People have not been paying enough attention to fairness metrics for recommender systems as well. For accuracy performance, people have proposed MAE, precision@K, and nDCG, among many other metrics. For fairness, there has been no consensus agreement on which metric should be used. Hao Wang [5] proposes a statistical concept called Degree of Matthew Effect. In this paper, we conduct experiments evaluated by Degree of Matthew Effect and MAE in the Experiment Section.

Information geometry [6] arises as a popular research topic in recent days. One important building stone of information geometry is the idea of divergence. Divergence has been introduced as a new distance measure. It has been used to design an optimization technique called natural gradient learning. However, in 2020, researchers [7] prove that natural gradient learning is equivalent to Gauss-Newton Method.

In this paper, we borrow the concept Kullback-Leibler Divergance from Information Geometry and build a fair recommender algorithm with regularization based on this concept. In the Experiment Section, we demonstrate our algorithm is more robust than vanilla matrix factorization and outperforms it by a large margin evaluated by fairness metrics.

## II. Related Work

Recommender system is one of the most widely adopted technologies in the internet industry. Due to its flexibility and profitability, researchers and industrial practitioners have developed algorithms such as SVD++ [8], Alternating Least Squares [9] and DeepFM [10], and many other techniques [11][12][13].

The very first batch of recommender systems such as collaborative filtering [14] and linear models such as logistic regression [15] and matrix factorization [8][9][10] mostly focus on optimizing RMSE of MAE scores of the user rating matrix. Starting from early 2010's, learning to rank technologies have emerged as a main research theme in the area. Learning to rank techniques such as Bayesian Personalized Ranking [16] and CLiMF [17] are popular apppproaches in commercial environments.

Fairness problems of recommender systems have become a hot research topic since 2017. In the year of 2017, Ed Chi [1] from Google Research proposes a fairness framework named Focused Learning, which takes advantage of regularization approach to reduce the unfairness of recommender systems. Other researchers resort to regularization for Learning to Rank paradigm to resolve the issue. In 2021, Hao Wang [5] introduces a mathematically solid regularization technique named Zipf Matrix Factorization that attacks the fairness problem with a statistical estimator based regularization term.

### III. INFORMATION GEOMETRY

Information Geometry is a scientific field invented by Japanese researcher Amari. The major building stone of Information Geometry is the concept of divergence. The formal definition of divergence is as follows [6] :

**Definition 3.1** $D[P:Q]$ is called a divergence when it satisfies the following criteria:
(1) $D[P:Q] \geq 0$.
(2) $D[P:Q] = 0$, when and only when $P = Q$.
(3) When $P$ and $Q$ are sufficiently close, by denoting their coordinates by $\xi_P$ and $\xi_Q = \xi_P + d\xi$, the Taylor expansion of $D$ is written as

$$D[\xi_P : \xi_P + d\xi] = \frac{1}{2} g_{ij}(\xi_P) d\xi_i d\xi_j + O(|d\xi|^3),$$

and matrix $\mathbf{G} = g_{ij}$ is positive-defifinite, depending on $\xi_P$.

Divergence is a function defined on 2 points on a manifold and the function is anti-symmetric, namely $D[P:Q] \neq D[Q:P]$. By definition, divergence has many concrete examples. If we assume there are 2 probability distributions $p(x)$ and $q(x)$ defined on manifold M, the Kullback-Leibler Divergence is defined as follows :

$$D_{KL}[p(x):q(x)] = \int p(x) \log \frac{p(x)}{q(x)} dx$$

The discrete version of Kullback-Leibler Divergence is defined in the following formula :

$$D_{KL}[p(x):q(x)] = \sum p(x) \log \frac{p(x)}{q(x)}$$

Kullback-Leibler Divergence is a suitable measure to compute the discrepancies between probability distributions.

Consider the context setting of recommendation, the most "fair" distribution of output item data structure would be each item consists of $\frac{1}{n}$ of the total item list. Therefore the distance between the recommendation output structure and the uniform distribution serves as an ideal measure for fairness of the algorithm. The distance is naturally defined as the Kullback-Leibler Divergence.

### IV. KL-MAT : REGULARIZATION VIA INFORMATION GEOMETRY

Matrix Factorization is one of the most successful recommender system algorithms in the past decades. The regularized matrix factorization framework is defined to optimize the following loss function :

$$\sum_{i=1}^{m} \sum_{j=1}^{n} \left( \frac{R_{ij}}{R_{max}} - \frac{U_i^T \cdot V_j}{||U_i|| \cdot ||V_j||} \right)^2 + \beta F(\cdot)$$

, where $R_{ij}$ is the rating score user i gives item j, $R_{max}$ is the maximum value of all rating scores. $U_i$ is the user feature vector and $V_j$ is the item feature vector. $\beta$ is the regularization coefficient, and $F(\cdot)$ is the regularization term.

We take advantage of the regularized matrix factorization scheme and KL-divergence and defines the loss function of our newly invented recommender system algorithm as follows:

$$L = \sum_{i=1}^{m} \sum_{j=1}^{n} \left( \frac{R_{ij}}{R_{max}} - \frac{U_i^T \cdot V_j}{||U_i|| \cdot ||V_j||} \right)^2 + \beta \left( \sum_{j=1}^{n} p\left(\frac{1}{rank_j}\right) \log \left( \frac{p(rank_j)}{q(rank_j)} \right) + \sum_{j=1}^{n} q\left(\frac{1}{rank_j}\right) \log \left( \frac{q(rank_j)}{p(rank_j)} \right) \right)$$

, where $R_{ij}$ is the rating score user i gives item j, $R_{max}$ is the maximum value of all rating scores. $rank_j$ is the popularity rank of item j in the output.

In the loss function L, $p(rank_j)$ is the distribution of item popularity ranks, while $q(rank_j)$ is uniform distribution, which means :

$$q(rank_j) = \frac{1}{n}$$

Kullback-Leibler Divergence is asymmetric, so we symmetricize it in the following way, as in our loss function :

$$E = D_{KL}[p(x):q(x)] + D_{KL}[q(x):p(x)]$$

Since $rank_j$ is unknown before computation, we approximate the value using linear combinations of dot products of user feature vectors and item feature vectors :

$$rank_j = \sum_{i=1}^{m} \alpha_i U_i^T \cdot V_j$$

We could model the user feature vectors and item feature vectors as the same as the ones in the loss function L. Now the question is how to compute the values of $\alpha_i$.

We first approximate the values of $rank_j$ by vanilla matrix factorization and at the same time use the user feature vectors and item feature vectors obtained to compute the values of $\alpha_i$ using Lasso Regression with positive coefficients. After we compute the values of $\alpha_i$, we could plug in its values as constants together with user feature vectors and item feature vectors as variables into the loss funciton L and solve L using Stochastic Gradient Descent.

In the problem setting of Stochastic Gradient Descent, we are to compute the gradient of the following loss function :

$$L' = \left(\frac{R_{ij}}{R_{max}} - \frac{U_i^T \cdot V_j}{||U_i|| \cdot ||V_j||}\right)^2 + \beta\left(\frac{1}{\alpha_i U_i^T \cdot V_j} - \frac{1}{n}\right)\left(\ln\left(\frac{1}{\alpha_i U_i^T \cdot V_j}\right) - \ln\left(\frac{1}{n}\right)\right)$$

The loss funtion $L'$ can be solved by standard Stochastic Gradient Descent as follows :

$$\frac{\partial L'}{\partial U_i} = -\left(\frac{2\left(\frac{R_{ij}}{R_{max}} - \frac{t_4}{t_2}\right)}{t_2}V_j - \frac{2 \cdot t_3 \cdot \left(\frac{R_{ij}}{R_{max}} - \frac{t_3}{t_2}\right)}{t_1 \cdot t_0^3}U_i + \frac{\beta(\ln(t_5) - \ln(t_6))}{t_7}V_j + \frac{\beta(\ln(t_5) - \ln(t_6))}{t_5 \cdot t_7}V_j\right)$$

, where :

$$t_0 = ||U_i||_2$$
$$t_1 = ||V_j||_2$$
$$t_2 = t_0 \cdot t_1$$
$$t_3 = V_j' \cdot U_i$$
$$t_4 = U_i' \cdot V_j$$

$$t_5 = \frac{1}{\alpha_i \cdot t_4}$$

$$t_6 = \frac{1}{n}$$

$$t_7 = \alpha_i \cdot t_4^2$$

, and :

$$\frac{\partial L'}{\partial V_j} = -\left(\frac{t_4}{t_2}U_i - \frac{t_3 \cdot t_4}{t_0 \cdot t_1^3}V_j + \frac{\beta(\ln(t_5) - \ln(t_6))}{t_7}U_i + \frac{\beta(\ln(t_5) - \ln(t_6))}{t_5 \cdot t_7}U_i\right)$$

, where :

$$t_0 = ||U_i||_2$$
$$t_1 = ||V_j||_2$$
$$t_2 = t_0 \cdot t_1$$
$$t_3 = V_j' \cdot U_i$$

$$t_4 = 2\left(\frac{R_{ij}}{R_{max}} - \frac{t_3}{t_2}\right)$$

$$t_5 = \frac{1}{\alpha_i \cdot t_3}$$

$$t_6 = \frac{1}{n}$$

$$t_7 = \alpha_i \cdot t_3^2$$

In the next section, we conduct experiments to compare the performance between KL-Mat and Vanilla Matrix Factorization. For the algorithmic accuracy evaluation, we use MAE score, which is defined as follows :

$$\text{MAE} = |R_{ij} - U_i^T \cdot V_j|$$

For the fairness metric, we resort to Degree of Matthew Effect as introduced by Hao Wang in [5]. Degree of Matthew Effect assumes the item frequency distribution of the algorithmic output follows power law distribution, and an estimator is calculated to approximate the exponent of the hypothetical power law distribution. The formula for Degree of Matthew Effect is defined as follows :

$$s = 1 + n\left(\sum_{i=1}^{n} \ln\frac{\text{rank}_i}{\text{rank}_{max}}\right)^{-1}$$

, where $rank_i$ is the popularity rank of the i-th item in the output. Since Degree of Matthew Effect is an approximation to estimate the exponent of the power law distribution of item popularity ranks, it is the deterministic variable to estimate the gravity of unfairness of the algorithmic output data structures.

## V. EXPERIMENT

We compare KL-Mat with Vanilla Matrix Factorization on 2 datasets, namely MovieLens Small Dataset [18] and MovieLens 1 Million Dataset [18]. The MovieLens Small Dataset consists of 610 users and 9724 items. Fig.1 demonstrates that our algorithm is more robust than Vanilla Matrix Factorization evaluated by MAE metric. However, at certain sub-interval of the regularization coefficient parameter spectrum, our algorithm is inferior to Vanilla Matrix Factorization.

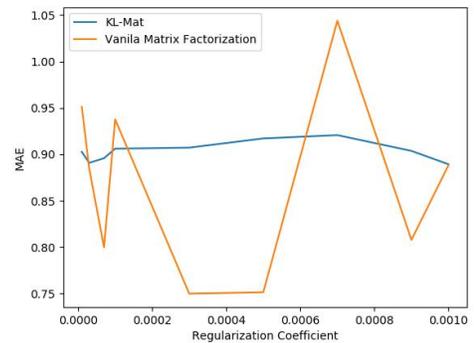

Fig. 1 Comparison between KL-Mat and Vanilla Matrix Factorization on MAE score (MovieLens Small Dataset)

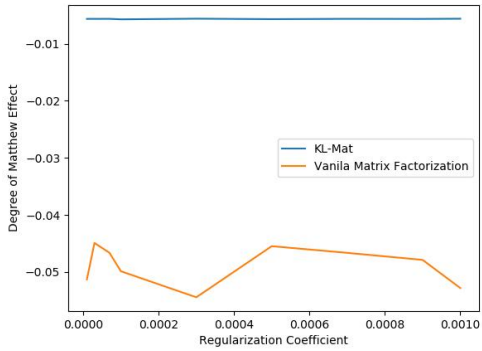

Fig. 2 Comparison between KL-Mat and Vanilla Matrix Factorization on Degree of Matthew Effect (MovieLens Small Dataset)

From Fig. 2 we observe that KL-Mat has far superior performance than Vanilla Matrix Factorization when compared by Degree of Matthew Effect. Based on our first experiment on MovieLens Small Dataset, we are safe to draw the conclusion that there exists a trade-off between algorithmic accuracy and fairness for KL-Mat. Obviously, the tendency to flatten out the item popularity rank distribution in the output structures does not help improve the accuracy performance.

MovieLens 1 Million Dataset consists of 6040 users and 3706 items. For this dataset, KL-Mat performs better than itself on MovieLens Small Dataset (Fig. 3). Once again, KL-Mat is more robust than Vanilla Matrix Factorization with only a small spread of variance on the data set.

Fig. 4 demonstrates the superiority of KL-Mat over Vanilla Matrix Factorization in fairness metrics. The performance of KL-Mat is also more stable than Vanilla Matrix Factorization.

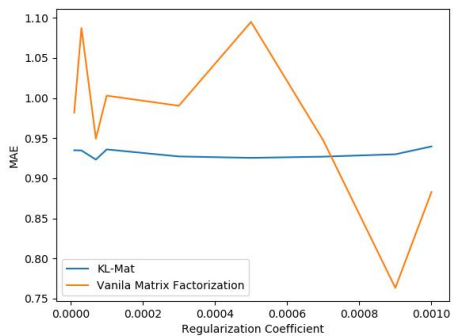

Fig. 3 Comparison between KL-Mat and Vanilla Matrix Factorization on MAE score (MovieLens 1 Million Dataset)

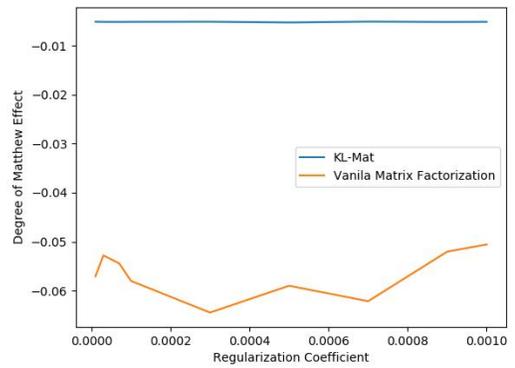

Fig. 4 Comparison between KL-Mat and Vanilla Matrix Factorization on Degree of Matthew Effect (MovieLens 1 Million Dataset)

## VI. CONCLUSION

In this paper, we propose a novel regularization technique for recommender system named KL-Mat. KL-Mat takes the form of regularized matrix factorization and outperforms Vanilla Matrix Factorization by a large margin by fairness metrics. Although by accuracy metrics, KL-Mat is inferior to Vanilla Matrix Factorization, it is more robust with much lower variance in MAE performance.

In future work, we would like to understand the trade-off between accuracy and fairness by KL-Mat. We would also like to investigate whether uniformly distributed popularity ranks should be the distribution in the regularization term.